\documentclass{svproc}
\usepackage{url}
\usepackage{graphicx}

\usepackage{hyperref}

\begin{document}
\mainmatter              
%
\title{Localization and classification of intracranial hemorrhages in CT data}
\titlerunning{ICH localization and classification}  
%
\author{Jakub Nemcek\inst{1} \and Roman Jakubicek\inst{1} \and Jiri Chmelik\inst{1} }
\authorrunning{Jakub Nemcek et al.} 
%
\tocauthor{Jakub Nemcek, Roman Jakubicek and Jiri Chmelik}
\institute{Department of Biomedical Engineering, Faculty of Electrical Engineering and Communication, Brno University of Technology, Brno, CZE\\
\email{xnemce04@vutbr.cz},\\ WWW home page:
\texttt{\url{https://www.vutbr.cz/en/people/jakub-nemcek-185952}} \\ 
\textbf{PREPRINT - submitted on EMBEC 2020 - paper has not been reviewed yet} 
}

\maketitle              

\begin{abstract}
Intracranial hemorrhages (ICHs) are life-threatening brain injures with a relatively high incidence. In this paper, the automatic algorithm for the detection and classification of ICHs, including localization, is present. The set of binary convolutional neural network-based classifiers with a designed cascade-parallel architecture is used. This automatic system may lead to a distinct decrease in the diagnostic process's duration in acute cases. An average Jaccard coefficient of 53.7~\% is achieved on the data from the publicly available head CT dataset CQ500.

\keywords{intracranial hemorrhage, convolutional neural network, localization, classification, computed tomography}
\end{abstract}
\section{Introduction}
Intracranial hemorrhages (ICHs) are life-threatening brain injures with a relatively high incidence (25 cases per 100,000 persons per year), which can be caused by physical trauma or non-traumatically (hemorrhagic stroke) \cite{Caceres2012}. There are five elementary sub-types of ICHs -- intraparenchymal (IPH), intraventricular (IVH), subdural (SDH), epidural (EDH) and subarachnoid hemorrhages (SAH). The early diagnosis of the ICH is important to plan the appropriate therapy. In contrast, quick treatment is crucial to reduce the risk of the negative (in many cases, permanent) consequences, even death.

The most frequently used ICH diagnostic method is an x-ray computed tomography (CT); however, it produces 3D image data that needs to be examined and described by a radiologist in the shortest possible time. Nowadays, computer-aided diagnosis (CAD) systems became a helpful tool, which can help to reduce required examination time and/or to prevent the omissions (e.g., small or indistinct pathology).

Deep learning, especially convolutional neural networks (CNN), proved to be one of the strongest state-of-the-art methods used for image analysis, including many medical fields \cite{Ker2018,Litjens2017,Shen20170621}. Several approaches using CNNs and their modifications for analysis of ICH in brain CT data has been recently published \cite{articleSegm}. Some authors aimed to a binary classification of individual axial slices (presence of hemorrhage) only. In these cases, a single 2D VGG16 CNN \cite{Castro2019} or 2D CNN in combination with a bi-directional LSTM \cite{Patel2019} was used. In other papers, the approaches was extended by the classification of ICH into the sub-types, such as in \cite{KARKI2020,Lee2019}, where simple 2D classification CNN was used. The authors in \cite{Nguyen2020} proposed the pre-trained ResNet-like CNNs combined with bi-directional LSTM for the classification purpose. Some authors solved more precise localization of ICH than the axial slice position only. It is done by segmentation with 2D U-Net improved by affinity graphs \cite{ChoJungrae2019}, but they did not deal with ICH sub-type classification. A few papers, where authors solved all the above-described problems (presence of hemorrhage, classification of sub-type, precise localization/segmentation) was published. A simple, fully convolutional network (FCN) was used in \cite{Kuo2019}. More complex approaches using a cascade of classification CNN (presence of hemorrhage in a slice) and subsequent multi-class segmentation FCN \cite{ChoJunghwan2019}, or 3D joint CNN and recurrent neural network for the classification modified by Grad-CAM for a precise position estimation \cite{Ye2019} have been recently presented.

In this paper, the automatic algorithm for the detection and classification of ICHs, including also localization, is present. The set of binary CNN-based classifiers with a designed cascade-parallel architecture enables the localization of ICHs even in the case of the combination of sub-types in a single scan. Moreover, the proposed algorithm provides the result in the form of labeled 3D bounding boxes (BBs), the determination of which is based on a fusion of partial detections from a set of CNNs utilizing three 2D orthogonal slices. Such an automatic system thus may especially lead to a distinct decrease in the duration of the diagnostic process in acute cases, which in case of manual detection by a radiologist can take more than 5 minutes.

\section{Methods}
\begin{figure}	[!tb]
			\centering
			\includegraphics[width=0.95\linewidth]{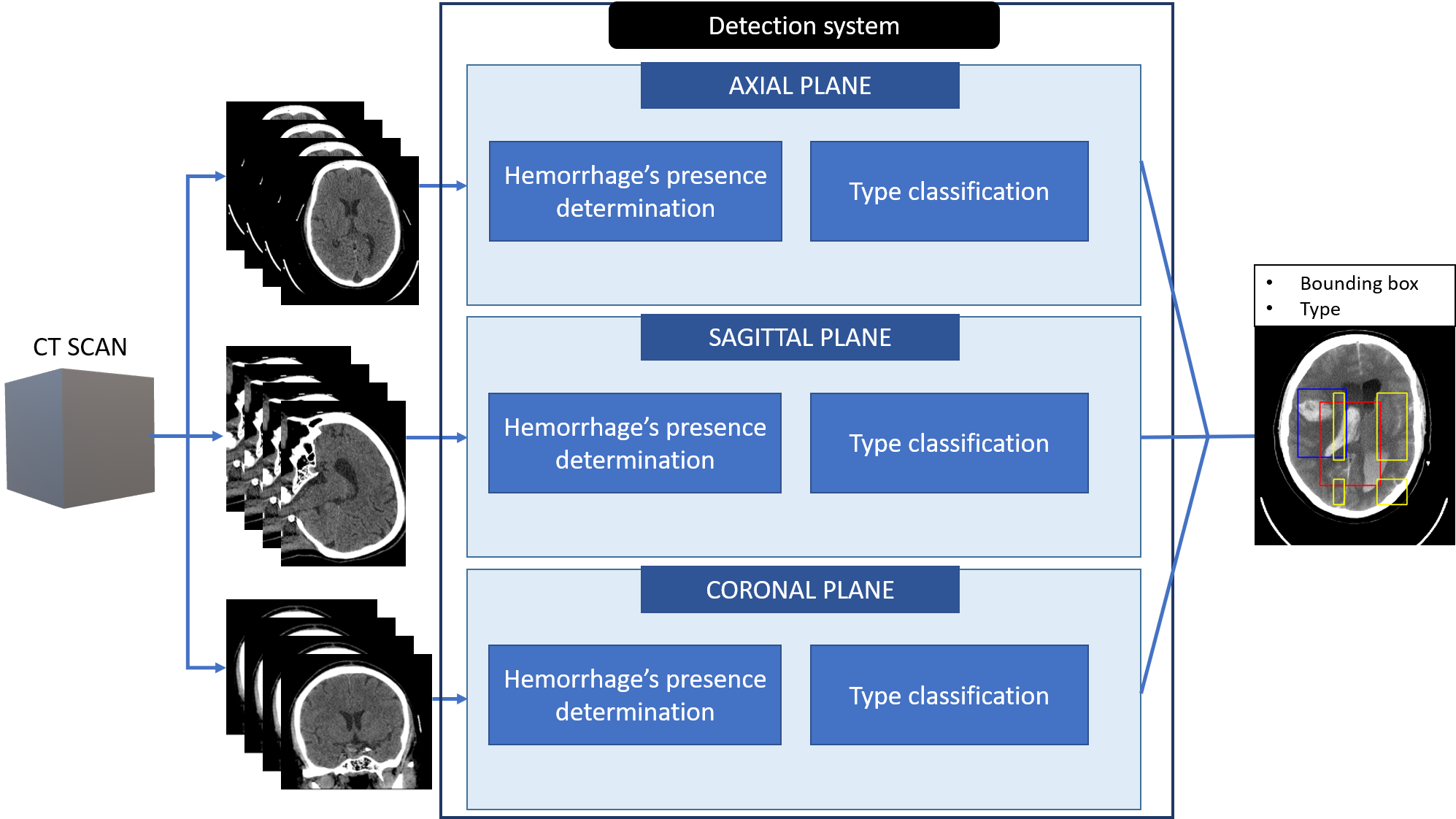}
			\caption{Block scheme of the three subsystems of the proposed approach.}
		\label{fig:bScheme_whole}
\end{figure}

The algorithm for localizing and classifying the type of the ICHs is based on orthogonal 2D CT slices analysis. Three independent subsystems based on a series of classification CNN models are proposed, as shown in Fig.~\ref{fig:bScheme_whole}. Together, the subsystems for individual slices of the CT scan in mutually perpendicular anatomical planes (axial, sagittal, and coronal) provide the final 3D rectangular BBs, and the classification of hemorrhages occurred in the CT scan.

\subsection{Experimental data}
 
The data from the publicly available head CT dataset CQ500 \cite{HeadStudy2018} was used in this work (194 CT scans with and 221 without a hemorrhagic finding). In the case of positive injured, the database contains all of the five sub-types of hemorrhages, including also annotations in the form of patient diagnose, which are based on the evaluation by three independent radiologists. However, the amount of data with different sub-types is highly imbalanced and in the case of EDH is insufficient (13 scans). For our purposes, the expert slice-level annotations are required. Therefore, the original annotations were manually expanded with ICH localizations in the way of labeling the occurrence and the sub-type of hemorrhages in the individual slices of CT scans in the axial, sagittal, and coronal plane.

\subsection{Pre-processing}
For our purpose, only non-contrast data is used, and in the first step, it is rotationally aligned by the algorithm published in \cite{Chmelik2019}. Further, the individual slices in axial, sagittal, and coronal planes are re-sampled into the space with a slice thickness of 5 mm along the Z-axis. 

According to \cite{HeadStudy2018}, the three contrast-enhanced images with special radiological windows (i.e., brain, subdural, and bone) are used; it forms three-channel input into the classifiers of the subsystems.

\subsection{Classifier design} 
The task of the first subsystem's CNN classifier is to determine whether there is a hemorrhage that occurred in the slice. In the case of positive output, the slice is processed by the series of four binary CNN classifiers which determine the presence or absence of the particular type of hemorrhage in the slice with exception of EDH due to it's insufficient incidence in our dataset. In each scan, final BB for every hemorrhage's type is given as an intersection of the partial outputs.

For the classifiers mentioned above, a CNN model is trained to predict the probabilities of two categories. To include 3D information the final class of a slice is derived from the weighted fusion of output probabilities, which are also predicted for adjacent parallel slices.

\subsection{Implementation details}
The proposed algorithm was implemented in Matlab. For each of the classifiers, the CNN architecture Inception-ResNet-v2 \cite{IncResNet2016} pre-trained on the ImageNet dataset \cite{ImageNet2009} was chosen and fine-tuned to be able to fulfil particular task. All networks were trained for eight epochs with the batch-size of 24 using weighted cross-entropy as the loss function, L2-regularization (regularization factor is set as 0.0005) and Adam \cite{Kingma2015AdamAM} optimizer with an initial learning rate of 0.0001 ($\beta_1=0.9$ and $\beta_2=0.999$). The slices are resized to the size of $299 \times 299$ pixels. 
Augmentation of the training data (70~\% of scans) is applied in the form of slight translation, rotation and scaling, and random reflection. The number of considered adjacent slices to predict a final class is two or four (depending on the classifier).

\section{Results and discussion}


Thirty percent of the CT scans were randomly selected for the evaluation of the final algorithm. The Jaccard coefficient (also known as Intersection over Union -- IoU) has been chosen as an evaluation metric, which expresses the relative overlap of ground truth and detected BB. 
In order to determine the sensitivity (Se) and positive predictive value (PPV), the determination of the number of true positive (TP), false positive (FP), and false negative (FN) detections is needed. It is usually performed by using thresholding of IoU. The thresh value has been chosen 0.5~IoU to take into account the properties of this metric in 3D space, especially the effect of object size on the interpretation of results.

\begin{table}[!htb]
\centering
\caption{Achieved results of proposed algorithm on the testing dataset. The IoU value reflects relative overlap between ground truth and detected bounding box. The sensitivity (Se) and positive predictive value (PPV) is calculate with a thresh value of 0.5~IoU. The bold values in the last column present the weighted average through all of ICH types.}
\label{iou}
\begin{tabular}{l||c|c|c|c||c}
    &  \textbf{\ IPH\ } & \textbf{\ IVH \ } & \textbf{\ SDH\ }  & \textbf{\ SAH\ } & \textbf{Total ICHs}\\ \hline
\hline
\textbf{$\#$ cases}         & 40    &  8    & 18    & 16   &  82      \\ \hline
\textbf{IoU [\%]}           & 59.0  & 53.5  & 43.2  & 52.5 & \bfseries 53.7    \\ \hline
\textbf{Se$_{0.5}$ [\%]}    & 62.5  & 62.5  & 38.9  & 50.0 & \bfseries 54.9    \\ \hline
\textbf{PPV$_{0.5}$ [\%]}   & 80.7  & 83.3  & 77.8  & 61.5 & \bfseries 76.6    \\ \hline

\end{tabular}
\end{table}

\begin{figure}	[!tb]
			\centering
			\includegraphics[width=0.95\linewidth]{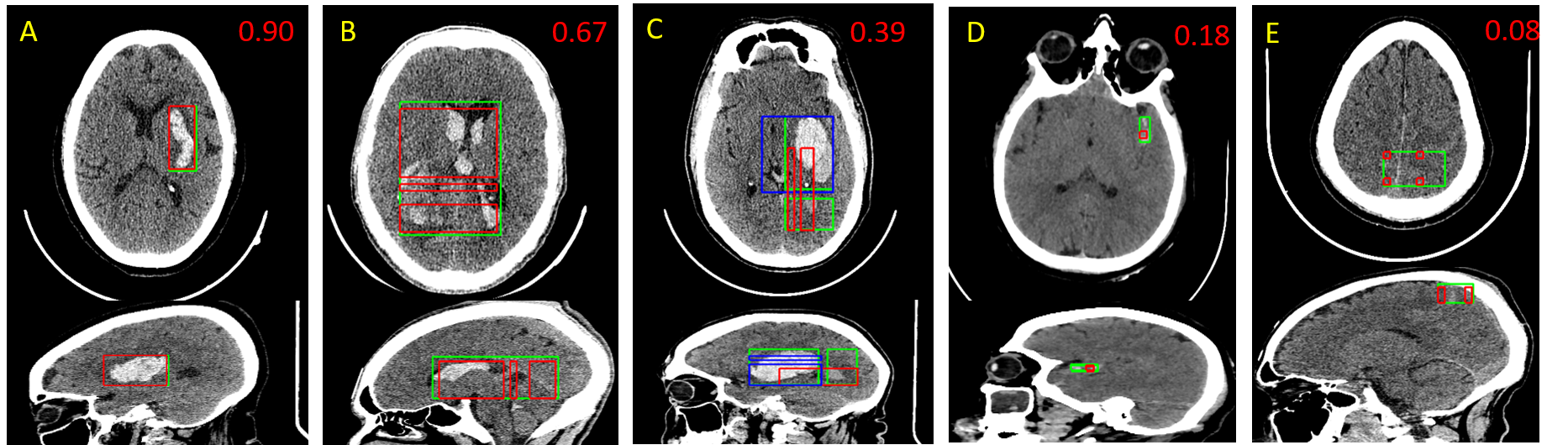}
			\caption{Axial and sagittal views of resulting bounding boxes of the algorithm testing compared to ground truth (green) together with the IoU value (in the right upper corner). A -- bounding box of IPH (red) matching well with ground truth, B -- result of IVH localisation (red), C -- localisation of multiple hemorrhage types: IPH (blue) and IVH (red) with the total IoU for both types jointly, D -- small SAH localization (red), E -- poorly located SAH with small overlap between the found bounding box and ground truth.}
		\label{fig:bboxesIoU}
\end{figure}

The IoU value can be low for tiny detected objects, although by subjective evaluation it seems to be a correct localization (Fig.~\ref{fig:bboxesIoU}~D); the slight difference in the position of BBs causes a distinct decrease in their relative overlap.

Based on achieved sensitivities (Table~\ref{iou}) it is evident that the algorithm provides more false-negative cases; thus the ICHs are not detected or, more frequently, were detected inaccurately (Fig.~\ref{fig:bboxesIoU}~D~E). It could have been caused by an unbalanced number of positive and negative images in the training dataset (negative cases were up to five times more frequent than positive ones). In contrast, it caused a lower number of false-positive cases and higher PPV.

The CNN models can distinguish various types of haemorrhages despite their different position, size and shape variability, hence the proposed algorithm can localize the ICH and provide its label even in case of small hemorrhages (see Fig.~\ref{fig:bboxesIoU}~D). 

Despite small amount of data with IVH, the results are similar to other types, which may be caused by their characteristic position and shape (see Fig.~\ref{fig:bboxesIoU}~B), which is given by the ventricles of brain. Lower results of SDH may be caused by the problematic detection in sagittal and coronal direction. In addition, it is observed that EDHs may sometimes be confused with SDHs (which may be a problem even for a human observer, mainly in sagittal and coronal slices).

The main benefit of the proposed architecture is locating and classifying ICHs, even in cases of the appearance of multiple sub-types of ICHs (Fig.~\ref{fig:bboxesIoU}~C). 
There can be predicted more FN cases if the image contains small ICH and large ICH of another type. 

The total duration of the algorithm is about 2 minutes per patient, which may lead to earlier hemorrhage identification. Moreover, the great potential for computer-aid-diagnostic systems is also given by the method's ability to warn a radiologist, roughly delimit and classify the affected region, and hence minimize the chances for missing the hemorrhage by oversight. It may help to prevent permanent disability or even death.


\section{Conclusion}
This paper presents an automatic algorithm for the localization and classification of ICHs, that provides labeled 3D rectangular bounding boxes as an output. The proposed method is based on a set of binary CNN classifiers with a designed cascade-parallel architecture, that enables the localisation and classification of the ICHs even in CT scans with combined types of ICHs. An average Jaccard coefficient of 53.7~\% is achieved on data from the publicly available dataset.

\section*{Acknowledgement}
The Titan Xp GPU used for this research was donated by the NVIDIA Corporation.
\section*{Conflict of Interest}
Authors declare none.
%
%
%
\bibliographystyle{splncs03}
\bibliography{mybibfile}

\end{document}